\documentclass[nopubldata]{lpor2014}
\begin{document}
%%%
% The following  four lines will be completed by the publisher:
% \setpages[]{}
% \setvolume[]{}
% \setyear{}%
% \setdoi{}%
%%%
%
% Here, the file name of the title figure is to be given in the form
% \titlefigure{<filename>}.
% If you don't want a figure here, just omit or comment out this line.
\titlefigure{Fig1}

% Give an abstract text:
\abstract{
Quantum ground-state cooling of macroscopic mechanical resonators is of
essential importance to both fundamental physics and applied science.
Conventional method of laser cooling is limited by the quantum backaction,
which requires mechanical sideband resolved in order to cool to ground state.
This work presents an idea to break the quantum backaction limit by engineering
intracavity optical squeezing. It gives rise to quantum interference for all the dissipation channels,
and under certain circumstances can totally remove the influence of the cavity dissipation and the
resultant quantum backaction, with much lower cooling limit
irrespective of the sideband resolution.
We show that our scheme enables ground-state cooling in the highly unresolved sideband limit and it
also works beyond the weak coupling regime,
which provides the opportunity for quantum manipulation of macroscopic mechanical systems.
}

% In the following lines, title and authoring informaton are to be
% given.  First the article's main title:
\title{Intracavity-squeezed optomechanical cooling}
%
% If the article has a very long title or lots of authors, then
% please give shorter versions of that for the column title:
\titlerunning{}
% Please name all authors and tag them with their respective
% institute(s) using the \inst directive.  The corresponding author
% gets an additional asterisk, e.g., as in
% \author{First Author\inst{1,*}, Second Author\inst{2,3}, and Third Author\inst{2}}
% If there is just one affiliation the numbers are omitted.
\author{Jing-Hui Gan\inst{1}, Yong-Chun Liu\inst{1,*}, Cuicui Lu\inst{2}, Xiao Wang\inst{1}, Meng Khoon Tey\inst{1}, Li You\inst{1}}
%
% Provide a short name list for the column title.  If there are more
% than two authors, please indicate that in the form "F. Author et
% al."  Here are the schemes for one, two, and three or more authors,
% respectively:
% \authorrunning{F. Author}
% \authorrunning{F. Author and S. Author}
\authorrunning{Jing-Hui Gan et al.}
%
% Here the institute(s) should be given.  If there is more than one,
% the entries are separated by \and tags.
% \institute{%
%   University of ...
% \and
%   University of ...
% }
\institute{%
State Key Laboratory of Low-Dimensional Quantum Physics, Department
of Physics, Frontier Science Center for Quantum Information, Collaborative Innovation Center of Quantum Matter, Tsinghua University, Beijing 100084, P. R. China

\and
Beijing Key Laboratory of Nanophotonics and Ultrafine Optoelectronic Systems, School of Physics,
 Beijing Institute of Technology, Beijing 100081, P. R. China
}
%
% Give the e-mail address of the corresponding author:
\mail{\email{ycliu@tsinghua.edu.cn}}
%
% Give some keywords for the article, if you have:
%
%%%
% The following  four lines will be completed by the publisher:
% \received{}
% \revised{}
% \accepted{}
% \published{\block}
%%%
\maketitle
% Here, the main text starts.

\section{Introduction}
Quantum manipulation of macroscopic objects is a persistent goal in quantum
science and related application
fields \cite{bocko1996measurement,knobel2003nanometre}. An essential prerequisite to observe quantum
phenomenon concerns cooling, i.e., reducing the random thermal motional energy.
In the past half century, the development of various cooling methods have revolutionized the field of atomic physics \cite{phillips1998nobel,wineland2013nobel}.
In recent years, the study of macroscopic mechanical resonators in the quantum regime \cite{kippenberg2008cavity,aspelmeyer2014cavity} has emerged as an important new frontier with
extensive potential applications in quantum information \cite{o2010quantum,verhagen2012quantum,wang2012using,dong2012optomechanical},
quantum-limited measurement \cite{lahaye2004approaching,teufel2009nanomechanical,McClelland2011,krause2012high,purdy2013observation,bbli2018mag}
and fundamental test of quantum mechanics \cite{romero2011large,pepper2012optomechanical,blencowe2013effective,sekatski2014macroscopic}.
Ground-state cooling of mechanical resonators can be realized using cavity optomechanical systems with a laser driving the red mechanical sideband \cite{wilson2007theory,marquardt2007quantum}, in the analogous sprit of laser cooling of neutral atoms \cite{hansch1975cooling} and ions \cite{abcd}.
 Such sideband cooling scheme has been adopted in various optomechanical experiments with great successes \cite{gigan2006self,arcizet2006radiation,rocheleau2010preparation,teufel2011sideband,chan2011laser,peterson2016laser}.

However, quantum backaction, a pure quantum effect originating from the counter-rotating-wave interaction between optical photons and mechanical phonons, can cause serious heating of the mechanical resonator.
It erects a quantum backaction limit corresponding to a minimum achievable phonon occupancy inversely related to the sideband resolution $\omega_{m}/\kappa$, where $\omega_{m}$ is the mechanical resonance frequency and $\kappa$ is the cavity dissipation rate. To cool below this quantum backaction limit, new schemes are needed. Sub-quantum-backaction cooling schemes are especially important for
optomechanical systems operating in the unresolved sideband limit (approximately $4\omega_{m}/\kappa<1$), where ground-state cooling is impossible with conventional sideband cooling scheme \cite{wilson2007theory,marquardt2007quantum}. For example, macroscopic mechanical resonators typically possess low resonance frequencies, preventing from reaching the resolved sideband limit. Recently, several proposals suggested better cooling performance, relying on ideas such as pulsed driving \cite{machnes2012pulsed,wang2011ultraefficient}, dissipative coupling \cite{elste2009quantum,li2009reactive,xuereb2011dissipative}, squeezed driving \cite{clark2017sideband}, feedback controlled light\cite{rossi2017enhancing,zippilli2018cavity} as well as hybrid approaches \cite{liu2015coupled,gu2013quantum,guo2014electromagnetically,liu2015optomechanically,ojanen2014ground,chen2015cooling,genes2009micromechanical}. However, complete removal of the quantum backaction heating due to cavity dissipation is not realized. For instance, the effect of the intrinsic cavity dissipation $\kappa_0$ cannot be eliminated using the typical noise interference approach \cite{weiss2013quantum}, while additional dissipation from the ancillary components takes effect in the hybrid approaches \cite{liu2015coupled}.

\begin{figure}[b]
\centering
\includegraphics[width=0.9\linewidth]{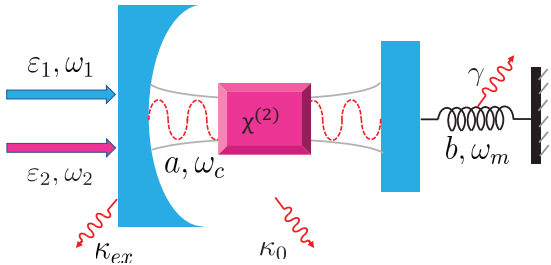}\caption{(Color online) A
schematic description of a cavity optomechanical system with a second-order
nonlinear medium in the cavity. The cavity is driven by a driving laser and
the nonlinear medium is driven by a nonlinear pumping laser.}%
\label{fig:model}%
\end{figure}

Here we proposed an intracavity-squeezed optomechanical cooling scheme to
completely remove the influence of cavity dissipations from all
channels. By putting a second-order ($\chi^{(2)}$) nonlinear medium inside the cavity
optomechanical system, the intracavity field can be strongly squeezed, which facilitates quantum noise interference for all the dissipation channels including both the
external and intrinsic cavity dissipations. As a result, the quantum backaction heating effect
associated with the cavity dissipations can be fully removed, the resulting cooling
limit is no longer dependent upon the sideband resolution.
Therefore, the scheme we introduce enables ground-state cooling
with full removal of the resolved sideband
restriction.
Moreover, it works even when the
noise interference model becomes invalid beyond the weak coupling regime.
The analytical results we obtain show that
the final cooling limit only depends on the $Q_{m}/n_{\mathrm{th}}$ ratio, where
$Q_{m}$ is the mechanical quality factor and $n_{\mathrm{th}}$ is the ambient thermal
phonon number.

\section{The Model}
The system we consider includes a $\chi^{(2)}$ nonlinear medium inside
an optomechanical cavity, as depicted in Fig.~\ref{fig:model}. In the
frame rotating at the driving laser frequency $\omega_{p}$, the Hamiltonian of
the system is given by%
\begin{align}
H &  =\omega_c a_1^{\dag}a_1+2\omega_ca_2^{\dag} a_2+\omega_{m}b^{\dag}b+(\nu a_1^2a_2^\dag+\nu^{\ast}a_1^{\dag 2}a_2)
\nonumber\\
&  +g_{1}a_1^{\dag} a_1(b+b^{\dag})+g_{2}a_2^{\dag} a_2(b+b^{\dag})+H_{\textrm{drive}},
\end{align}
where $a_1$ and $a_2$ are the annihilation
operators of the optical fundamental mode and second-order mode, and $\nu$ is the single photon $\chi^{(2)}$ nonlinearity, $b$ is the mechanical mode.
$g_1 (g_2)$ is the corresponding single photon optomechanical coupling between optical modes and mechanical mode. $H_{\textrm{drive}}=(\varepsilon_1 e^{i\omega_1 t}a_1+\varepsilon_2 e^{i\omega_2 t}a_2+H.C.)  $ is the laser driving term, where $\varepsilon_1 (\omega_1)$ is the driving strength (frequency) of optical fundamental mode and $\varepsilon_2 (\omega_2)$ is the pumping magnitude (frequency) of second-order optical mode.

Similar model has been studied for modifying the normal mode
splitting \cite{huang2009normal}, enhancing the effective photon-phonon
interaction \cite{lu2015squeezed} and improving position detection
sensitivity \cite{peano2015intracavity}. Experimentally, such a model is
readily implemented using a Fabry-P\'{e}rot cavity or a whispering-gallery
mode cavity \cite{ding2010high,furst2011quantum}.

Within stability regime, the optical and mechanical
modes reach steady states denoted by $\alpha_{1,2}=\langle a_{1,2}\rangle_{\mathrm{ss}}$
and $\beta=\langle b\rangle_{\mathrm{ss}}$. After carrying out a linearization
procedure, we can write down the Langevin equations of the three mode (see section VI in the Supporting information). It will be found that apart from the needed parametric amplifier term induced by the nonlinear medium, the effect induced by the radiation pressure of the second-order optical mode should be noted. We found that the radiation pressure of the second order mode will introduce, firstly a modification of the detuning and optomchanical coupling of the fundamental optical mode which is not essential. Secondly added noise and effect can be safely neglected when $\Delta_{2,eff}\gg\textrm{max}[\sqrt{g_{2}^2|\alpha _{2}|^{2}\kappa _{2}/\omega_m},\sqrt{\kappa_2/\kappa_1}|\nu\alpha_1|]$, where $\Delta_{2,eff}=\omega_2-2\omega_c-g_2(\beta+\beta^{\ast})$ is the effective detuning of the second-order optical mode. The detailed calculation can be found in supporting information and similar demonstration of neglecting the effect of the second-order optical mode for displacement measurement can be found in ref\cite{peano2015intracavity}. With this approximation the system Hamiltonian reduces to $H_{L}=-\Delta
a_{1}^{\dag}a_{1}+\omega_{m}b_{1}^{\dag}b_{1}+G(a_{1}+a_{1}^{\dag}%
)(b_{1}+b_{1}^{\dag})+(\varepsilon a_{1}^{\dag2}+\varepsilon^{\ast}a_{1}^{2})$, where
$a_{1}=a-\alpha$ and $b_{1}=b-\beta$ represents quantum fluctuations
around the steady states, $\Delta$ is the effective detuning
and $G$ is the modified optomechanical coupling
strength (assumed to be real without loss of generality) and $\varepsilon=\nu\alpha_2$ denotes the squeezing parameter.  The linearized
quantum Langevin equations are given by%
\begin{align}
\dot{a_{1}} &  =(i\Delta-\frac{\kappa}{2})a_{1}-iG(b_{1}+b_{1}^{\dag
})-2i\varepsilon a_{1}^{\dag}-\sqrt{\kappa}a_{\mathrm{in}},\label{eq:langevin}\\
\dot{b_{1}} &  =(-i\omega_{m}-\frac{\gamma}{2})b_{1}-iG(a_{1}+a_{1}^{\dag
})-\sqrt{\gamma}b_{\mathrm{in}},
\end{align}
where $\kappa$ represents the total cavity dissipation including both the
external dissipation $\kappa_{\mathrm{ex}}$ and the intrinsic dissipation
$\kappa_{0}$, $\gamma=\omega_{m}/Q_{m}$ is the mechanical dissipation rate,
$a_{\mathrm{in}}$ and $b_{\mathrm{in}}$ are the noise operators satisfying
$\langle a_{\mathrm{in}}(t)a_{\mathrm{in}}^{\dag}(t^{^{\prime}})\rangle
=\delta(t-t^{^{\prime}})$, $\langle b_{\mathrm{in}}(t)b_{\mathrm{in}}^{\dag
}(t^{^{\prime}})\rangle=(n_{\mathrm{th}}+1)\delta(t-t^{^{\prime}})$, $\langle
b_{\mathrm{in}}^{\dag}(t)b_{\mathrm{in}}(t^{^{\prime}})\rangle=n_{\mathrm{th}%
}\delta(t-t^{^{\prime}})$, and $n_{\mathrm{th}}=1/(e^{\hbar\omega_{m}/k_{B}%
T}-1)$ is the ambient thermal phonon number with $T$ being the corresponding ambient temperature.

\section{Weak coupling Regime}
In the weak coupling regime with $G\ll(\kappa,\omega_{m})$, the optomechanical
cooling effect is determined by the spectrum of the optical force $S_{FF}%
(\omega)=\int d\tau e^{i\omega\tau}\langle F(\tau)F(0)\rangle$, where
$F=G(a_{1}^{\dag}+a_{1})/x_{\mathrm{ZPF}}\propto X_1$ is the optical force with $X_1$ being the quadrature of the optical field and $x_{\mathrm{ZPF}}$ being the
zero point fluctuation of the mechanical mode. According to Fermi's golden
rule, the cooling and heating rates are given by $\Gamma_{-}=S_{FF}(\omega
_{m})x_{\mathrm{ZPF}}^{2}$ and $\Gamma_{+}=S_{FF}(-\omega_{m})x_{\mathrm{ZPF}%
}^{2}$, corresponding to the ability to absorb and emit a phonon by the
cavity, respectively. These rates then determine the net cooling rate
$\Gamma_{\mathrm{opt}}=\Gamma_{-}-\Gamma_{+}$ and the effective phonon number
$n_{\mathrm{opt}}=\Gamma_{+}/(\Gamma_{-}-\Gamma_{+})$. Using the above
relations, we obtain (section III in the Supporting Information)
\begin{align}
S_{FF}(\omega) &  =\frac{{G}^{2}\kappa}{x_{\mathrm{ZPF}}^{2}}\left\vert
\chi({\omega})\frac{1+{2i\varepsilon}^{\ast}\chi^{\ast}(-{\omega})}{1-4\left\vert
{\varepsilon}\right\vert ^{2}\chi({\omega})\chi^{\ast}(-{\omega})}\right\vert
^{2},\label{SFF}\\
\Gamma_{\mathrm{opt}} &  =\frac{-4G^{2}\kappa\omega_{m}(\Delta+2\varepsilon_{r}%
)}{(\kappa^{2}/4+\Delta^{2}-4|\varepsilon|^{2}-\omega_{m}^{2})^{2}+\omega_{m}^{2}%
\kappa^{2}},\label{Gammaopt}\\
n_{\mathrm{opt}} &  =\frac{(\omega_{m}+\Delta+2\varepsilon_{r})^{2}+(\kappa
/2+2\varepsilon_{i})^{2}}{-4\omega_{m}(\Delta+2\varepsilon_{r})}.\label{nopt}
\end{align}
where $\varepsilon_{r}$ ($\varepsilon_{i}$)$\ $is the real (imaginary) part of ${\varepsilon}$, and
$\chi(\omega)=[\kappa/2-i(\omega+\Delta)]^{-1}$ is the optical response
function. The final phonon occupancy in the weak coupling regime is
then given by $n_{f}^{\mathrm{wk}}=n_{\mathrm{opt}}+\gamma n_{\mathrm{th}%
}/\Gamma_{\mathrm{opt}}$.

Without the nonlinear pumping ($\varepsilon=0$), the above results reduce to the
conventional sideband cooling case
\cite{wilson2007theory,marquardt2007quantum}, with $S_{FF}^{\varepsilon=0}(\omega
)={G}^{2}\kappa\left\vert \chi({\omega})\right\vert ^{2}/x_{\mathrm{ZPF}}^{2}%
$, and $n_{\mathrm{opt}}^{\varepsilon=0}=[(\omega_{m}+\Delta)^{2}+\kappa^{2}%
/4]/(-4\omega_{m}\Delta)$. In the unresolved sideband regime, both cooling
rate and heating rate stay almost the same, leading to very small net cooling
rate and thus the cooling is inefficient. The minimum effective phonon number
obtained at the optimal detuning $\Delta=-\kappa/2$ is given by
$n_{\mathrm{opt}}^{\varepsilon=0}=\kappa/(4\omega_{m})$, which is strongly limited by
the sideband resolution, and ground-state cooling is unattainable in the unresolved sideband limit.

In the scheme we present, from Eq. (\ref{SFF}) we can find that the noise
interference effect appears as the result of the nonlinear pumping.
To suppress the
quantum backaction heating, destructive interference should occur at $\omega=-\omega_{m}$.
By setting $S_{FF}(-\omega_{m})=0$, we obtain the optimal condition
\begin{equation}
{\varepsilon=}\frac{1}{2i\chi({\omega}_{m})}=-\frac{{\omega}_{m}{+\Delta}}{2}%
-i\frac{\kappa}{4}.\label{S}%
\end{equation}
In this case the quantum backaction heating process is completely cancelled ($\Gamma_{+}=0$),
and the optical mode behaviours as a zero-temperature bath ($n_{\mathrm{opt}}=0$).
Therefore, the final phonon number is no longer
limited by the sideband resolution anymore, and ground-state cooling is
attainable for arbitrary large $\kappa/\omega_{m}$.

The unique advantage of our intracavity squeezing scheme is that the improved
cooling performance is immune to all the cavity dissipation channels, including both
the external dissipation $\kappa_{\mathrm{ex}}$ and intrinsic dissipation $\kappa_{0}$.
This is because the intracavity field is squeezed with an internal nonlinear process,
which results in the complete suppression of the quantum fluctuations of the intracavity field for certain quadrature component ($X_1$ in our case).
Because the squeezing does not depend on the input-output process, the noise interference takes place for
all the dissipation channels.
This property is essentially different from the
squeezed driving outside the cavity \cite{clark2017sideband}, where the noise interference occurs on
the input-output boundary, and the noise associated with the intracavity
dissipation $\kappa_{0}$ cannot interfere with the squeezed input light
\cite{walls2007quantum}, with final phonon occupancy still limited by
$\kappa_{0}/(4\omega_{m})$.

\begin{figure}[t]
\centering
\includegraphics[width=0.95\linewidth]{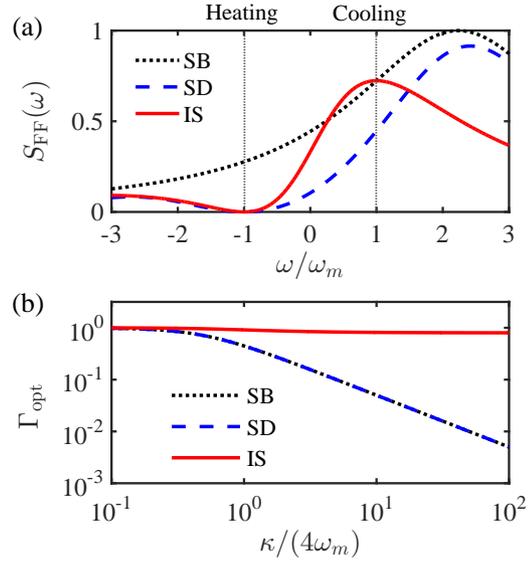}
\caption{
(Color online) (a) Normalized noise power spectrum of the backaction force $S_{FF}(\omega)$ in three schemes: sideband cooling scheme (SB, black dotted curve), squeezed driving scheme (SD, blue dashed curve), and intracavity squeezing scheme (IS, red solid curve). The two
vertical gray solid lines corresponds to frequencies $\omega=\pm\omega_m$, which determines the cooling and heating rates, respectively. The normalization factor is $4G^{2}/(\kappa x_{\mathrm{ZPF}}^{2})$. (b) Normalized net cooling rates $\Gamma_{\mathrm{opt}}$ versus the ratio $\kappa/(4\omega_m)$ for the three schemes. The normalization factor is $4G^{2}/\kappa$.
Other unspecified parameters are $\kappa/(4\omega_m)=1$, $\Delta=-\sqrt{\kappa^2/4+\omega_m^2}$.
}%
\label{fig:ps}%
\end{figure}

Moreover, at the optimal condition given by Eq. (\ref{S}), we obtain
$\Gamma_{-}={G}^{2}\kappa\left\vert \chi(\omega
_{m})\right\vert ^{2}=\Gamma_{-}^{\varepsilon=0}$, indicating that the cooling process is totally unaffected
compared with the conventional sideband cooling case.
This result is also quite
different from that of the squeezed driving scheme.
For the latter, the suppression of heating process
is accompanied by a reduction of cooling rate $\Gamma_{-}$, with unchanged net cooling
rate $\Gamma_{\mathrm{opt}}$ compared with the conventional sideband cooling scheme.
The comparisons under typical parameters are plotted in Figs.~\ref{fig:ps}(a) and \ref{fig:ps}(b).
They show that the net cooling rate in our intracavity squeezing scheme is $\kappa/(4\omega_{m})$
times larger than that of both sideband cooling and squeezed driving schemes.
This enhancement factor is especially important in the unresolved sideband limit with $\kappa/(4\omega_{m})\gg1$.

%\begin{figure}[t]
%\centering
%\includegraphics[width=0.9\linewidth]{Fig3.eps}\caption{(Color
%online) Time evolution of the phonon occupancy $n(t)$ for the three different
%schemes with $\kappa/(4\omega_{m})=100$, $n_{\mathrm{th}}=10^{3}$, $Q_{m}=10^{5}$ and $\Delta=-\kappa/2$.
%The coupling strengths $G$ are optimized to achieve their respective best cooling limits for the three schemes: $G=8\omega_m$ for SD scheme, $G=5.3\omega_m$ for SB and IS schemes.}%
%\label{fig3}
%\end{figure}

\section{General results}

To obtain the general results of the cooling limits and cooling
dynamics beyond the weak coupling regime, we employ the covariance matrix
method by solving the time evolution of the mean phonon number
\cite{liu2013dynamic} based on the master equation $\dot{\rho}=i[\rho
,H_{L}]+\kappa\mathcal{D}[a_{1}]\rho+\gamma(n_{th}+1)\mathcal{D}[b_{1}%
]\rho+\gamma n_{th}\mathcal{D}[b_{1}^{\dag}]\rho$, where $\mathcal{D}[\hat
{o}]\rho=\hat{o}\rho\hat{o}^{\dag}-(\hat{o}^{\dag}\hat{o}\rho+\rho\hat
{o}^{\dag}\hat{o})/2$ denotes the Liouvillian in Lindblad form for operator
$\hat{o}$.
%In Fig.~\ref{fig3} we plot the cooling dynamics for different cooling
%schemes, which verifies that our intracavity squeezing approach is indeed
%advantageous for much larger cooling rate and much lower cooling limit.

The general steady-state cooling limit is obtained as%
\begin{align}
n_{f} &  =n_{f}^{\mathrm{wk}}+n_{f}^{\mathrm{st}},\label{nf}\\
n_{f}^{\mathrm{wk}} &  =\frac{\gamma n_{\mathrm{th}}}{\Gamma_{\mathrm{opt}}%
}=\frac{4(\Delta+\omega_{m})^{2}+\kappa^{2}}{4G^{2}\kappa}\gamma
n_{\mathrm{th}},\\
n_{f}^{\mathrm{st}} &  =\frac{G^{2}\left(  \frac{2\Delta+\omega_{m}}%
{\omega_{m}\kappa}\gamma n_{\mathrm{th}}-\frac{1}{2}\right)  }{\left(
2\Delta+\omega_{m}\right)  \omega_{m}+4G^{2}}+\frac{\gamma n_{\mathrm{th}}%
}{\kappa},\label{nfst}%
\end{align}
where we have used the large cooperativity assumption $C=4G^{2}/(\kappa\gamma)\gg1$, and used the optimal nonlinear pumping condition given by Eq.(\ref{S}). The latter is reasonable because we focus on the highly unresolved regime with $\kappa\gg G$, and the nonlinear pumping strength is on the same order of $\kappa$. The modification of optimal $\varepsilon$ caused by $G$ will be much smaller than the order of $\kappa$, thus we can still use Eq. (\ref{S}) for the optimal pumping strength beyond the weak coupling regime, which is also compatible with the numerical results. Here $n_{f}^{\mathrm{wk}}$ corresponds exactly to the result in the weak
coupling regime with $\Gamma_{\mathrm{opt}}$ given by Eq. (\ref{nopt}), while $n_{f}^{\mathrm{st}}$
represents the cooling limit originating from the strong coupling effect.
In Eq. (\ref{nfst}), the first term of $n_{f}^{\mathrm{st}}$ takes effect when $G$ is comparable
with $\sqrt{-\Delta\omega_{m}/2}$, and this term also implies the stability
condition $\left(  2\Delta+\omega_{m}\right)  \omega_{m}+4G^{2}<0$, i.e.,
$\Delta<-\omega_{m}/2$ and $G<\sqrt{-\left(  2\Delta+\omega_{m}\right)
\omega_{m}}/2$, which agree well with the result obtained using the
Routh-Hurwtiz method \cite{ghobadi2011quantum}; the second term $\gamma n_{\mathrm{th}}/{\kappa}$
can be neglected in the unresolved sideband regime since it scales inversely proportional to $\kappa$.

\begin{figure}[t]
\centering
\includegraphics[width=0.99\linewidth]{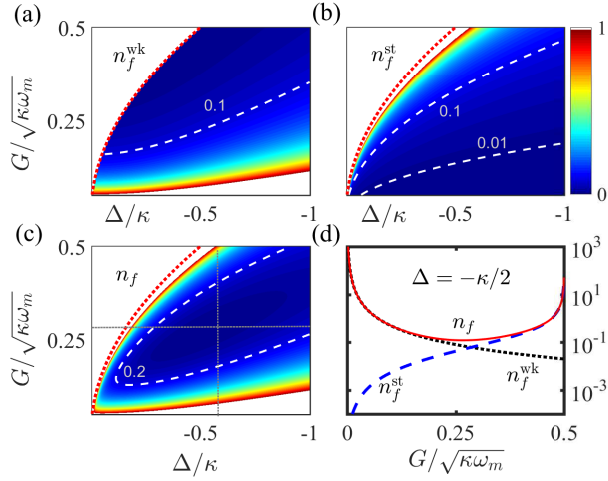}\caption{(Color online) Cooling limits
$n_{f}^{\mathrm{wk}}$ (a), $n_{f}^{\mathrm{st}}$ (b) and $n_{f}=n_{f}^{\mathrm{wk}}+n_{f}^{\mathrm{st}}$ (c) versus the detuning $\Delta$ and the
coupling strength $G$. The red dash-dotted curves
corresponding to the critical boundary between the stable and unstable regions. In (c), the vertical and horizontal lines corresponds
to the optimal choice of $\kappa$ and $G$. (d) $n_{f}^{\mathrm{wk}}$, $n_{f}^{\mathrm{st}}$ and $n_{f}$ versus $G$ for $\Delta=-\kappa/2$. Other parameters are $\kappa/(4\omega_{m})=100$, $n_{\mathrm{th}}=10^{3}$ and $Q_{m}=10^{5}$.}%
\label{fig4}
\end{figure}

Within the weak coupling picture, the increase of the coupling strength $G$ leads
to the increase of the cooling rate $\Gamma_{\mathrm{opt}}$ and thereby
reduces the cooling limit, as shown in Fig.~\ref{fig4}(a). However, when $G$ is strong enough, the strong
coupling effect and the stability condition constrains the cooling limit.
At the same time, the maximum achievable $G$ depends on the detuning $\Delta$, where a larger
detuning will tolerate a larger coupling strength [Fig.~\ref{fig4}(b)]. Therefore, both the detuning
$\Delta$ and the coupling strength $G$ should be optimized to obtain the
minimum cooling limit. The results are shown in Fig.~\ref{fig4}(c-d).
When $Q_{m}/n_{\mathrm{th}}\gg1$ and $|\Delta|\gg\omega_{m}$, we obtain the optimal detuning, coupling strength, and the
corresponding cooling limit as
\begin{gather}
\Delta=-\frac{\kappa}{2},\text{ }G=\sqrt{\frac{\kappa\omega_{m}}{4+\sqrt
{Q_{m}/n_{\mathrm{th}}}}},\\
n_{f}^{\min}=2\frac{n_{\mathrm{th}}}{Q_{m}}+\sqrt{\frac{n_{\mathrm{th}}}%
{Q_{m}}}.\label{nfmin}
\end{gather}
The above results indicates that: (1) the optimal detuning $\Delta=-\kappa/2$ is the same as the conventional sideband cooling scheme in
the unresolved sideband regime; (2) the maximum achievable $G$ scales as $\sqrt{\kappa\omega_m}$ and thus the maximum net cooling rate $\Gamma_{\mathrm{opt}}$ scales as $\omega_{m}$, which
ensures high cooling efficiency even when $\kappa$ is large; (3) the achievable cooling limit
only depends on the $Q_{m}/n_{\mathrm{th}}$ ratio, and the cavity dissipation
$\kappa$ still does not come into play, even in this general case.
From Eq. (\ref{nfmin}) we can derived the ground-state cooling condition as $Q_{m}/n_{\mathrm{th}}\gtrsim4$, corresponding
to $Q_{m}\omega_{m}\gtrsim4k_{B}T/\hbar$.

\begin{figure}[t]
\centering
\includegraphics[width=0.85\linewidth]{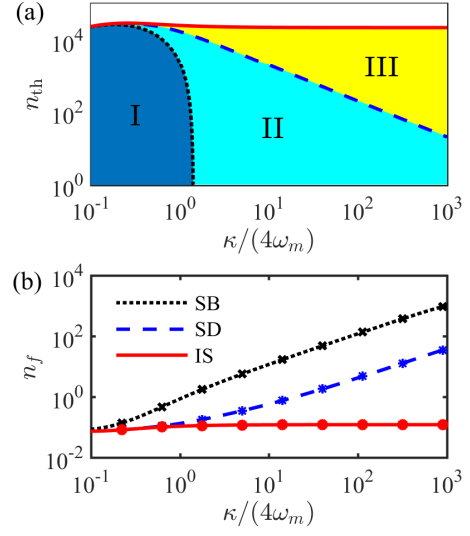}\caption{(Color
online) (a) Ground-state cooling regions versus $\kappa/4\omega_{m}$ and $n_{\mathrm{th}}$. The curves denote the boundary contours $n_{f}=1$ for the three schemes (numerical results). The regions where ground-state cooling is achievable for the three schemes are: SB--region I; SD--regions I and II; IS--regions I, II and III. %$\uppercase\expandafter{\romannumeral1}$
(b) Cooling limit of the three schemes for $n_{\mathrm{th}}=10^{3}$. Markers with red circles, blue stars and dark X-marks are numerical results and the curves are the corresponding analytical results.
In both (a) and (b) we have set $Q_{m}=10^{5}$, and other unspecified parameters are optimized to obtain the best cooling limits. }%
\label{fig:coolinglimit1}%
\end{figure}

For comparison, we also present the minimal cooling limits for the sideband cooling and
squeezed driving schemes (see section IV and V in the Supporting Information). For the sideband cooling scheme, $n_{f,\mathrm{SB}%
}^{\min}=\kappa/4\omega_{m}[1+n_{\mathrm{th}}/Q_{m}+2\sqrt{(n_{\mathrm{th}%
}/Q_{m})^{2}+n_{\mathrm{th}}/Q_{m}}]$, and ground-state cooling requires
$\kappa/(4\omega_{m})<1$; For squeezed driving scheme,
$n_{f,\mathrm{SD}}^{\min}=2c+\sqrt{c(1+4c)}$ with $c=\gamma n_{\mathrm{th}}%
\kappa/\omega_{m}^{2}$, and ground-state cooling requires $\kappa/(4\omega
_{m})<Q_{m}/(5n_{\mathrm{th}})$ (here we have neglected the $\kappa_0/(4\omega
_{m})$ term). Similar restriction also exists in the dissipative
coupling scheme \cite{weiss2013quantum}.
The parameter range of ground-state cooling regions for the three schemes are plotted in Fig.~\ref{fig:coolinglimit1}(a).
For the sideband cooling scheme, ground-state cooling is possible only in region I; For the squeezed driving scheme, both region I and II enables
ground-state cooling; while for our intracavity squeezing scheme, ground-state cooling can be realized in all the three regions I, II and III.
In Fig.~\ref{fig:coolinglimit1}(b), we
plot the achievable minimal final phonon number as a function of $\kappa/4\omega_{m}$ for $n_{th}=10^{3}$.
The above analytical results agree quite well with the numerical results.
It reveals that our intracavity squeezing scheme uniquely possesses the feature that the
cooling limit is independent of the cavity dissipation in the unresolved
sideband limit.

\section{Conclusion}
In summary, we have presented an intracavity squeezing scheme allowing
optomechanical cooling to the ground state below the quantum backaction limit. The
intracavity squeezing results in quantum noise destructive interference for all the dissipation channels, including the external and
intrinsic cavity dissipations. The quantum backaction heating is then strongly suppressed, leading to enhanced net cooling rate
and reduced cooling limit, enabling ground-state cooling under arbitrary
low sideband resolution. We derive
the cooling limit beyond the weak coupling regime and find that the final
cooling limit only depends on the $Q_{m}/n_{\mathrm{th}}$ ratio even when
$(\kappa,G)\gg\omega_{m}$, showing that ground-state cooling is attainable as
long as $Q_{m}/n_{\mathrm{th}}\gtrsim4$. This opens new possibilities for quantum manipulation of
massive and macroscopic mechanical systems with low resonance
frequencies.

\section*{Supporting Information}
Supporting information is available from the Wiley Online Library or from the author
% Give an acknowledgement if you have to thank someone, are
% supported by someone or something like that:
\begin{acknowledgement}
This work was supported by the National Natural Science Foundation of
China (NSFC) (Grants No. 91736106, 11674390, 91836302, 91850117, 11604378, 11654003 and 11654001), the National Key R\&D Program of China (Grant
2018YFA0306504 and 2018YFA0306503).
\end{acknowledgement}

Note added: we note that two related papers appeared on arXiv \cite{a1,a2} during the review process of our work.
% The authors' biographies are given inside an \authorbox command
% each, all together enclosed in this environment.  Here is the
% syntax:
% \authorbox{<name>}{Authorname}{Text}
% where "<name>" is the Filename of the portrait picture, if any; it
% should consist of the author's last name prefixed with "cv".
% "Authorname" is this respective author's name being bolded as the
% first word(s) of the biography text, the rest of which follows in
% the third argument.
%\begin{biographies}
%  \authorbox{cv}{}{}
%  \authorbox{cv}{}{}
%  \authorbox{cv}{}{}
%\end{biographies}
%
% Give your bibliography below.  If you are using BibTeX please make
% sure to enclose all .bib files needed!

%\end{document}
% Here, the article ends.

\clearpage
\onecolumn
\begin{center}
  \textbf{\large Supplementary Material}\\[.2cm]
%(Dated: \today)\\[1cm]
\end{center}

This Supplementary Material is organized as follows. In Sec. \ref{Sec1} we
present the linearization of the Hamiltonian and the quantum Langevin
equations. In Sec. \ref{Sec2} we analyze the stability conditions of the
system. In Sec. \ref{Sec3} we use the quantum noise approach to derive the
power spectrum of the radiation pressure in the weak coupling regime. The
general results of cooling limits are deduced in Sec. \ref{Sec4}. In Sec. %
\ref{Sec5} we compare the cooling limits of three different schemes. In Sec.%
\ref{Sec6} we discuss the noise from the second-order nonlinear mode.

\tableofcontents

\section{Linearized Hamiltonian}

\label{Sec1}
The hamitonian of the total system is

\begin{align}
H &  =\omega_c a_1^{\dag}a_1+2\omega_ca_2^{\dag} a_2+\omega_{m}b^{\dag}b+(\nu a_1^2a_2^\dag+\nu^{\ast}a_1^{\dag 2}a_2)
\nonumber\\
&  +g_{1}a_1^{\dag} a_1(b+b^{\dag})+g_{2}a_2^{\dag} a_2(b+b^{\dag})+H_{\textrm{drive}},
\end{align}
where $a_1$ and $a_2$ are the annihilation
operators of the optical fundamental mode and second-order mode, and $\nu$ is the single photon $\chi^{(2)}$ nonlinearity, $b$ is the mechanical mode.
$g_1 (g_2)$ is the corresponding single photon optomechanical coupling between optical modes and mechanical mode. $H_{\textrm{drive}}=(\epsilon_1 e^{i\omega_1 t}a_1+\epsilon_2 e^{i\omega_2 t}a_2+H.C.)  $ is the laser driving term, where $\epsilon_1 (\omega_1)$ is the driving strength (frequency) of optical fundamental mode and $\epsilon_2 (\omega_2)$ is the pumping magnitude (frequency) of second-order optical mode.

The effect of the second-order optical mode can be seen as classically with proper condition which will be discussed detailed in sec.~\ref{Sec6}. With this approximation, the reduced hamitonian yields
\begin{equation}
H=-\Delta _{0}a^{\dag }a+\omega _{m}b^{\dag }b+g_{0}a^{\dag }a(b+b^{\dag
})+(\Omega a^{\dag }+\Omega ^{\ast }a)+(\varepsilon a^{\dag
2}+\varepsilon^{\ast }a^{2}).
\end{equation}%
Here we work in a rotating frame with driving laser frequency $\omega _{p}$,
$a(a^{\dag })$ is the bosonic annihilation (creation) operator of the
optical mode with the angular resonance frequency $\omega _{c}$, $b(b^{\dag
})$ is the bosonic annihilation (creation) operator of the mechanical mode
with the angular resonance frequency $\omega _{m}$, $\Delta _{0}=\omega
_{p}-\omega _{c}$ is the driving laser detuning with respect to the cavity
resonance, $g_{0}$ is the single-photon optomechanical coupling strength, $%
\Omega $ ($\varepsilon$) is the driving (nonlinear pumping) strength with
laser frequency $\omega _{p}$ ($2\omega _{p}$). The quantum Langevin
equations of the system is
\begin{align}
\dot{a}& =(i\Delta _{0}-\frac{\kappa }{2})a-ig_{0}a(b+b^{\dag })-i\Omega
-2i\varepsilon a^{\dag }-\sqrt{\kappa _{\mathrm{ex}}}a_{\mathrm{ex,in}}-%
\sqrt{\kappa _{0}}a_{\mathrm{0,in}}, \\
\dot{b}& =(-i\omega _{m}-\frac{\gamma }{2})b-ig_{0}a^{\dag }a-\sqrt{\gamma }%
b_{\mathrm{in}},
\end{align}%
where $\kappa _{\mathrm{ex}}\ $($\kappa _{0}$) is cavity dissipation rate
from external (internal) channels and $\kappa =\kappa _{\mathrm{ex}}+\kappa
_{0}$ is the total dissipation rate, $\gamma $ is the mechanical dissipation
rate with the relation with mechanical quality factor $Q_{m}$ as $\gamma
=\omega _{m}/Q_{m}$, $a_{\mathrm{ex,in}},a_{\mathrm{0,in}}$ and $b_{\mathrm{%
in}}$ are the corresponding noise operators associated with the
dissipations, which satisfy $\langle a_{\mathrm{ex,in}}(t)a_{\mathrm{ex,in}%
}^{\dag }(t^{^{\prime }})\rangle =\langle a_{\mathrm{0,in}}(t)a_{\mathrm{0,in%
}}^{\dag }(t^{^{\prime }})\rangle =\delta (t-t^{^{\prime }})$, $\langle b_{%
\mathrm{in}}(t)b_{\mathrm{in}}^{\dag }(t^{^{\prime }})\rangle =(n_{\mathrm{th%
}}+1)\delta (t-t^{^{\prime }})$, $\langle b_{\mathrm{in}}^{\dag }(t)b_{%
\mathrm{in}}(t^{^{\prime }})\rangle =n_{\mathrm{th}}\delta (t-t^{^{\prime
}}) $ with other correlators vanish. The mean thermal phonon number $n_{%
\mathrm{th}}=1/(e^{\hbar \omega _{m}/k_{B}T}-1)\approx k_{B}T/(\hbar \omega
_{m})$, where $T$ is the environmental temperature and the second
approximate equality holds for $k_{B}T\gg \hbar \omega _{m}$. From the above
correlation functions it is easy to verify $\sqrt{\kappa }a_{\mathrm{in}}=%
\sqrt{\kappa _{\mathrm{ex}}}a_{\mathrm{ex,in}}+\sqrt{\kappa _{0}}a_{\mathrm{%
0,in}}$, where $a_{\mathrm{in}}$ represents the total optical noise operator
satisfying $\langle a_{\mathrm{in}}(t)a_{\mathrm{in}}^{\dag }(t^{^{\prime
}})\rangle =\delta (t-t^{^{\prime }})$.

When the system is in the steady state, we can perform the linearization
procedure by assuming $a=\alpha +a_{1}\ $and $b=\beta +b_{1}$, where $\alpha
=\langle a\rangle _{\mathrm{ss}}$ and $\beta =\langle b\rangle _{\mathrm{ss}%
} $ are the mean values of the optical and mechanical fields without
considering the quantum fluctuations, determined by
\begin{align}
0& =(i\Delta _{0}-\frac{\kappa }{2})\alpha -ig_{0}\alpha (\beta +\beta
^{\ast })-i\Omega -2i\varepsilon\alpha ^{\ast }, \\
0& =(-i\omega _{m}-\frac{\gamma }{2})\beta -ig_{0}|\alpha |^{2}.
\end{align}%
The redefined operators $a_{1}$ and $b_{1}$ represent the quantum
fluctuations around the mean values, with the corresponding quantum Langevin
equations
\begin{align}
\dot{a_{1}}& =(i\Delta -\frac{\kappa }{2})a_{1}-iG(b_{1}+b_{1}^{\dag
})-2i\varepsilon a_{1}^{\dag }-\sqrt{\kappa _{\mathrm{ex}}}a_{\mathrm{ex,in}%
}-\sqrt{\kappa _{0}}a_{\mathrm{0,in}}, \\
\dot{b_{1}}& =(-i\omega _{m}-\frac{\gamma }{2})b_{1}-i(Ga_{1}^{\dag
}+G^{\ast }a_{1}^{\dag })-\sqrt{\gamma }b_{\mathrm{in}},
\end{align}%
where we only keep the linear terms of the operators,$\ \Delta =\Delta
_{0}-g_{0}(\beta +\beta ^{\ast })$ is the modified detuning, $G=g_{0}\alpha $
is the linearized optomechanical coupling strength. Without loss of
generality, $\alpha $ is assumed to be real, which can be realized by
adjusting the initial phase of the driving laser. Thus $G$ is real and the
linearized system Hamiltonian can be inferred as
\begin{equation}
H_{L}=-\Delta a_{1}^{\dag }a_{1}+\omega _{m}b_{1}^{\dag
}b_{1}+G(a_{1}+a_{1}^{\dag })(b_{1}+b_{1}^{\dag })+(\varepsilon a_{1}^{\dag
2}+\varepsilon ^{\ast }a_{1}^{2}).
\end{equation}

\section{Stability conditions}

\label{Sec2} The dynamics of the fluctuations around the steady state can be
written as
\begin{equation}
\frac{d}{dt}\mathbf{V}=\mathbf{DV,}
\end{equation}%
where $\mathbf{V}=(a_{1},a_{1}^{\dag },b_{1},b_{1}^{\dag })^{T}$ and
\begin{equation}
\mathbf{D}=\left(
\begin{array}{cccc}
i\Delta -\frac{\kappa }{2} & -2i\varepsilon & -iG & -iG \\
2i\varepsilon^{\ast } & -i\Delta -\frac{\kappa }{2} & iG & iG \\
-iG & -iG & -i\omega _{m}-\frac{\gamma }{2} & 0 \\
iG & iG & 0 & i\omega _{m}-\frac{\gamma }{2}%
\end{array}%
\right) .
\end{equation}%
The stability condition requires that the eigenvalues of the dynamical
evolution matrix $D$ have no positive real part. By employing the
Routh-Hurwtiz method, the stability condition is obtained as
\begin{equation}
\frac{-4G^{2}\omega _{m}(\Delta +2\varepsilon_{r})}{(\kappa ^{2}/4+\Delta
^{2}-4|\varepsilon|^{2})(\omega _{m}^{2}+\gamma ^{2}/4)}<1
\end{equation}

Under the optimized condition with the real and imaginary parts of $%
\varepsilon$ given by $\varepsilon_{r}=(-\Delta -\omega _{m})/2$, $%
\varepsilon_{i}=-\kappa /4$ and with the assumption $Q_{m}\gg 1$, the
stability condition reduces to
\begin{align}
G& <\frac{\sqrt{-\left( 2\Delta +\omega _{m}\right) \omega _{m}}}{2}, \\
\Delta & <-\frac{\omega _{m}}{2}.
\end{align}

\section{Weak coupling regime}

\label{Sec3} In the frequency domain, the quantum Langevin equations are
given by
\begin{align}
-i\omega a_{1}(\omega )& =(i\Delta -\frac{\kappa }{2})a_{1}(\omega )-iG\left[
b_{1}(\omega )+b_{1}^{\dag }(\omega )\right] -2i\varepsilon a_{1}^{\dag
}(\omega )-\sqrt{\kappa }a_{\mathrm{in}}(\omega ),  \label{a1} \\
-i\omega b_{1}(\omega )& =(-i\omega _{m}-\frac{\gamma }{2})b_{1}(\omega )-iG%
\left[ a_{1}(\omega )+a_{1}^{\dag }(\omega )\right] -\sqrt{\gamma }b_{%
\mathrm{in}}(\omega ).
\end{align}%
In the weak coupling regime, we can neglect the effect of backaction by
setting $G=0$ in Eq. (\ref{a1}) and then obtain
\begin{align}
a_{1}(\omega )& =\sqrt{\kappa }\chi ({\omega })\frac{{-}\tilde{a}_{\mathrm{in%
}}{(\omega )+2i\varepsilon }\chi ^{\ast }(-{\omega })\tilde{a}_{\mathrm{in}%
}^{\dag }{(\omega )}}{1-4\left\vert {\varepsilon }\right\vert ^{2}\chi ({%
\omega })\chi ^{\ast }(-{\omega })}, \\
a_{1}^{\dag }(\omega )& =\sqrt{\kappa }\chi ^{\ast }(-{\omega })\frac{{%
-2i\varepsilon }^{\ast }\chi ({\omega })\tilde{a}_{\mathrm{in}}{(\omega ){-}}%
\tilde{a}_{\mathrm{in}}^{\dag }{(\omega )}}{1-4\left\vert {\varepsilon }%
\right\vert ^{2}\chi ({\omega })\chi ^{\ast }(-{\omega })}.
\end{align}%
where we have define the optical response function
\begin{equation}
\chi ({\omega })=\frac{1}{-i({\omega +\Delta )}+\frac{\kappa }{2}}.
\end{equation}%
Hence the radiation pressure force $F=-G(a_{1}^{\dag }+a_{1})/x_{\mathrm{ZPF}%
}$ can be expressed as
\begin{equation}
F{\left( {\omega }\right) }=\frac{{G}\sqrt{\kappa }}{x_{\mathrm{ZPF}}}\frac{%
\left[ 1+{2i\varepsilon }^{\ast }\chi ^{\ast }(-{\omega })\right] \chi ({%
\omega })\tilde{a}_{\mathrm{in}}{(\omega )+}\left[ 1-{2i\varepsilon }\chi ({%
\omega })\right] \chi ^{\ast }(-{\omega })\tilde{a}_{\mathrm{in}}^{\dag }{%
(\omega )}}{1-4\left\vert {\varepsilon }\right\vert ^{2}\chi ({\omega })\chi
^{\ast }(-{\omega })},
\end{equation}%
where $x_{\mathrm{ZPF}}=\sqrt{\hbar /(2m_{\mathrm{eff}}\omega _{m})}$ is the
zero-point fluctuation of the mechanical mode with $m_{\mathrm{eff}}$ being
the effective mass of the mechanical mode. The power spectrum of the
radiation pressure is then obtained as
\begin{align}
S_{\mathrm{FF}}(\omega )& =\frac{{G}^{2}\kappa }{x_{\mathrm{ZPF}}^{2}}\left\vert \chi
({\omega })\frac{1+{2i\varepsilon }^{\ast }\chi ^{\ast }(-{\omega })}{%
1-4\left\vert {\varepsilon }\right\vert ^{2}\chi ({\omega })\chi ^{\ast }(-{%
\omega })}\right\vert ^{2}  \notag \\
& =\frac{G^{2}\kappa }{x_{\mathrm{ZPF}}^{2}}\frac{(\omega -\Delta
-2\varepsilon _{r})^{2}+(\kappa /2+2\varepsilon _{i})^{2}}{(\kappa
^{2}/4+\Delta ^{2}-4|\varepsilon |^{2}-\omega ^{2})^{2}+\omega ^{2}\kappa
^{2}},
\end{align}%
where $\varepsilon _{r}$ and $\varepsilon _{i}\ $are the real and imaginary
parts of ${\varepsilon }$, respectively.

The corresponding heating rate and cooling rate are given by $\Gamma
_{+}=S_{\mathrm{FF}}(-\omega _{m})x_{\mathrm{ZPF}}^{2}$ and $\Gamma
_{-}=S_{\mathrm{FF}}(\omega _{m})x_{\mathrm{ZPF}}^{2}$, respectively. Thus the net
cooling rate and the effective phonon number are given by
\begin{align}
\Gamma _{\text{\textrm{opt}}}& =\Gamma _{-}-\Gamma _{+}=\frac{-4G^{2}\kappa
\omega _{m}(\Delta +2\varepsilon_{r})}{(\kappa ^{2}/4+\Delta
^{2}-4|\varepsilon|^{2}-\omega ^{2})^{2}+\omega _{m}^{2}\kappa ^{2}}, \\
n_{\mathrm{opt}}& =\frac{\Gamma _{+}}{\Gamma _{-}-\Gamma _{+}}=\frac{(\omega
_{m}+\Delta +2\varepsilon_{r})^{2}+(\kappa /2+2\varepsilon_{i})^{2}}{%
-4\omega _{m}(\Delta +2\varepsilon_{r})}.
\end{align}
By setting $S_{\mathrm{FF}}(-\omega _{m})=0$ we obtain the condition when the heating
process can be fully canceled as
\begin{equation}
{\varepsilon=}\frac{1}{2i\chi ({\omega }_{m})}=-\frac{{\omega }_{m}{+\Delta }%
}{2}-i\frac{\kappa }{4}.
\end{equation}%
Under this condition, we obtain $\Gamma _{+}={G}^{2}\kappa \left\vert \chi
(\omega _{m})\right\vert ^{2}=\Gamma _{+}^{\varepsilon=0}$.

\section{General cooling limits}

\label{Sec4} With the Linearized Hamiltonian, the master equation is given
by
\begin{equation}
\dot{\rho}=i[\rho ,H_{L}]+\mathcal{L}\rho ,
\end{equation}%
where $\rho $ is the density matrix and $\mathcal{L}$ is the Lindband
super-operator with
\begin{eqnarray}
\mathcal{L}\rho &=&\frac{\kappa }{2}(2a_{1}\rho a_{1}^{\dag }-a_{1}^{\dag
}a_{1}\rho -\rho a_{1}^{\dag }a_{1})  \notag \\
&&+\frac{\gamma }{2}(n_{\mathrm{th}}+1)(2b_{1}\rho b_{1}^{\dag }-b_{1}^{\dag
}b_{1}\rho -\rho b_{1}^{\dag }b_{1})  \notag \\
&&+\frac{\gamma }{2}n_{\mathrm{th}}(2b_{1}^{\dag }\rho
b_{1}-b_{1}b_{1}^{\dag }\rho -\rho b_{1}b_{1}^{\dag }).
\end{eqnarray}%
The mean values of all the second-order moment operators of the system $%
(\langle a^{\dag }a\rangle ,$ $\langle b^{\dag }b\rangle ,$ $\langle a^{\dag
}b\rangle ,$ $\langle ab^{\dag }\rangle ,$ $\langle ab\rangle ,$ $\langle
a^{\dag }b^{\dag }\rangle ,$ $\langle a^{2}\rangle ,$ $\langle a^{\dag
2}\rangle ,$ $\langle b^{2}\rangle ,$ $\langle b^{\dag 2}\rangle )$ are
determined by the set of differential equations
\begin{align}
\frac{d\langle a^{\dag }a\rangle }{dt}& =-iG(\langle a^{\dag }b\rangle
-\langle ab^{\dag }\rangle +\langle a^{\dag }b^{\dag }\rangle -\langle
ab\rangle )-2i(\varepsilon\langle a^{\dag 2}\rangle -\varepsilon^{\ast
}\langle a^{2}\rangle )-\kappa \langle a^{\dag }a\rangle , \\
\frac{d\langle b^{\dag }b\rangle }{dt}& =-iG(-\langle a^{\dag }b\rangle
+\langle ab^{\dag }\rangle +\langle a^{\dag }b^{\dag }\rangle -\langle
ab\rangle )-\gamma \langle b^{\dag }b\rangle +\gamma n_{\mathrm{th}}, \\
\frac{d\langle a^{\dag }b\rangle }{dt}& =[-i(\Delta +\omega _{m})-\frac{%
\kappa +\gamma }{2}]\langle a^{\dag }b\rangle -iG(\langle a^{\dag }a\rangle
-\langle b^{\dag }b\rangle +\langle a^{\dag 2}\rangle -\langle b^{2}\rangle
)+2i\varepsilon^{\ast }\langle ab\rangle , \\
\frac{d\langle ab\rangle }{dt}& =[i(\Delta -\omega _{m})-\frac{\kappa
+\gamma }{2}]\langle ab\rangle -iG(\langle a^{\dag }a\rangle +\langle
b^{\dag }b\rangle +\langle a^{2}\rangle +\langle b^{2}\rangle
+1)-2i\varepsilon\langle a^{\dag }b\rangle , \\
\frac{d\langle a^{2}\rangle }{dt}& =(2i\Delta -\kappa )\langle a^{2}\rangle
-2iG(\langle ab\rangle +\langle ab^{\dag }\rangle )-2i\varepsilon(2\langle
a^{\dag }a\rangle +1), \\
\frac{d\langle b^{2}\rangle }{dt}& =(-2i\omega _{m}-\gamma )\langle
b^{2}\rangle -2iG(\langle a^{\dag }b\rangle +\langle ab\rangle ).
\end{align}%
Particularly, when $\varepsilon=0$, the above equations reduces to the
results for the sideband cooling case.

With the above equations, under the optimal choose of driving $%
\varepsilon_{r}=(-\Delta -\omega _{m})/2$, $\varepsilon_{i}=-\kappa /4$, and
under the approximation $C=4G^{2}/(\kappa \gamma )\gg 1$, we can determine
the general results of the final phonon occupancy beyond the weak coupling
limit as
\begin{gather}
n_{f}=\frac{\gamma n_{\mathrm{th}}}{\Gamma _{\mathrm{opt}}^{^{\prime }}}+%
\frac{G^{2}}{-2(4G^{2}+\omega _{m}^{2}+2\omega _{m}\Delta )},
\label{eq::coolinglimit} \\
\Gamma _{\mathrm{opt}}^{^{\prime }}=\frac{\Gamma _{\mathrm{opt}}\Gamma _{1}}{%
\Gamma _{\mathrm{opt}}+\Gamma _{1}}, \\
\Gamma _{\mathrm{opt}}=\frac{4G^{2}\kappa }{4(\Delta +\omega
_{m})^{2}+\kappa ^{2}}, \\
\Gamma _{1}=\frac{2\kappa \omega _{m}[4G^{2}+\omega _{m}(2\Delta +\omega
_{m})]}{(\omega _{m}^{2}+G^{2})(2\Delta +\omega _{m})+4G^{2}\omega _{m}},
\end{gather}%
where $\Gamma _{\mathrm{opt}}\ $is the net cooling rate in the weak coupling limit, $%
\Gamma _{\mathrm{opt}}^{^{\prime }}$ represents the effective net cooling rate beyond
the weak coupling limit. A proper choose of $\Delta $ and $G$ will results
in a minimal final phonon occupancy. To meet the stability command, the
achievable effective optomechanical coupling $G$ is bounded by a maximum
value $G_{\max }=\sqrt{(-\omega _{m}^{2}-2\Delta \omega _{m})/4}$. Define $%
\Delta =-y\kappa /2$ and $G=\sqrt{xy\kappa \omega _{m}/4}$, then generally
the minimal value of the final phonon number happens when $x=x^{\ast }$ and $%
y=2\sqrt{1-x}/(2-x)$, where $x^{\ast }$ is the solution in interval $(0,1)$
of the following equation
\begin{equation}
x^{4}+64(\frac{n_{\mathrm{th}}}{Q_{m}})^{2}(x-1)(x^{2}-6x+4)=0
\end{equation}%
Specially, when $n_{\mathrm{th}}/Q_{m}\ll 1$, $x^{\ast }=4\sqrt{n_{\mathrm{th%
}}/Q_{m}}/(4\sqrt{n_{\mathrm{th}}/Q_{m}}+1)$, and the minimal phonon number
is $n_{\min }=2n_{\mathrm{th}}/Q_{m}+\sqrt{n_{\mathrm{th}}/Q_{m}}$; when $n_{%
\mathrm{th}}/Q_{m}\gg 1$, then $x^{\ast }=3-\sqrt{5}$, and the minimum is $%
n_{\min }=\sqrt{(22+10\sqrt{5})/4}n_{\mathrm{th}}/Q_{m}$.

\section{Cooling limits of different schemes}

\label{Sec5}We analyze the cooling limit of three different schemes:
sideband cooling scheme, squeezed driving scheme and intracavity squeezing
scheme proposed here. In the following we focus on the unresolved sideband
regime. The general form of the final phonon occupancy can be described by
\begin{equation}
n_{f}=\frac{\gamma n_{\mathrm{th}}}{\Gamma _{\mathrm{opt}}^{^{\prime }}}%
+n_{\mathrm{opt}}^{^{\prime }}.
\end{equation}

For sideband cooling scheme, under the optimal detuning $\Delta =-\sqrt{%
\kappa ^{2}/4+\omega _{m}^{2}}$, it yields
\begin{align}
\Gamma _{\mathrm{opt}}^{^{\prime }}& =\frac{8G^{2}\omega _{m}^{2}(\kappa
\omega _{m}-4G^{2})}{\kappa (\kappa \omega _{m}-2G^{2})^{2}}, \\
n_{\mathrm{opt}}^{^{\prime }}& =\frac{\kappa \omega _{m}-2G^{2}}{\kappa
\omega _{m}-4G^{2}}\frac{\kappa }{4\omega _{m}}.
\end{align}%
The minimal value is obtained when $G=\sqrt{x\omega _{m}\kappa }$, where $%
x=1/(4+\sqrt{1+Q_{m}/n_{\mathrm{th}}})$ and the minimum phonon number is
\begin{equation}
n_{f,\mathrm{SB}}^{\min }=\frac{\kappa }{4\omega _{m}}(1+\frac{n_{\mathrm{th}%
}}{Q_{m}}+2\sqrt{\frac{n_{\mathrm{th}}}{Q_{m}}(\frac{n_{\mathrm{th}}}{Q_{m}}%
+1)}.
\end{equation}

For squeezed driving scheme, the differential equations are given by
\begin{align}
\frac{d\langle a^{\dag }a\rangle }{dt}& =-iG(\langle a^{\dag }b\rangle
-\langle ab^{\dag }\rangle +\langle a^{\dag }b^{\dag }\rangle -\langle
ab\rangle )+\kappa \sinh R-\kappa \langle a^{\dag }a\rangle , \\
\frac{d\langle b^{\dag }b\rangle }{dt}& =-iG(-\langle a^{\dag }b\rangle
+\langle ab^{\dag }\rangle +\langle a^{\dag }b^{\dag }\rangle -\langle
ab\rangle )-\gamma \langle b^{\dag }b\rangle +\gamma n_{\mathrm{th}}, \\
\frac{d\langle a^{\dag }b\rangle }{dt}& =[-i(\Delta +\omega _{m})-\frac{%
\kappa +\gamma }{2}]\langle a^{\dag }b\rangle -iG(\langle a^{\dag }a\rangle
-\langle b^{\dag }b\rangle +\langle a^{\dag 2}\rangle -\langle b^{2}\rangle
), \\
\frac{d\langle ab\rangle }{dt}& =[i(\Delta -\omega _{m})-\frac{\kappa
+\gamma }{2}]\langle ab\rangle -iG(\langle a^{\dag }a\rangle +\langle
b^{\dag }b\rangle +\langle a^{2}\rangle +\langle b^{2}\rangle +1), \\
\frac{d\langle a^{2}\rangle }{dt}& =(2i\Delta -\kappa )\langle a^{2}\rangle
-2iG(\langle ab\rangle +\langle ab^{\dag }\rangle )+\frac{1}{2}\kappa \sinh
(2R)e^{2i\phi }, \\
\frac{d\langle b^{2}\rangle }{dt}& =(-2i\omega _{m}-\gamma )\langle
b^{2}\rangle -2iG(\langle a^{\dag }b\rangle +\langle ab\rangle ),
\end{align}%
where $R$ is the squeezing magnitude and $\phi $ is the squeezing phase. The
optimal conditions are $\Delta =-\sqrt{\kappa ^{2}/4+\omega _{m}^{2}}$, $%
R=\sinh ^{-1}[\kappa /(2\omega _{m})]/2$ and $\phi =\arctan [-4\Delta \kappa
/(\kappa ^{2}+4\omega _{m}^{2}-4\Delta ^{2})]$. Under these optimal
conditions, the results are
\begin{align}
\Gamma _{\mathrm{opt}}^{^{\prime }}& =\frac{8G^{2}\omega _{m}^{2}(\kappa \omega
_{m}-4G^{2})}{\kappa (\kappa \omega _{m}-2G^{2})^{2}}, \\
n_{\mathrm{opt}}^{^{\prime }}& =\frac{G^{2}}{2(\kappa \omega -4G^{2})}.
\end{align}%
Define $c=\kappa n_{\mathrm{th}}/(4\omega _{m}Q_{m})$, then the minimal value is
obtained when $G=\sqrt{x\kappa \omega _{m}}$, where $x=c/(4c+\sqrt{c(1+4c)})$%
, and the minimum phonon number is
\begin{eqnarray}
n_{f,\mathrm{SD}}^{\min } &=&2c+\sqrt{c(1+4c)}  \notag \\
&=&\sqrt{\frac{\kappa n_{\mathrm{th}}}{4\omega _{m}Q_{m}}}\left( 2\sqrt{\frac{\kappa
n_{\mathrm{th}}}{4\omega _{m}Q_{m}}}+\sqrt{1+\frac{\kappa n_{\mathrm{th}}}{\omega _{m}Q_{m}}}%
\right)
\end{eqnarray}

For intracavity squeezing scheme, in the limit $n_{\mathrm{th}}\ll Q_{m}$, the
optimal choice of detuning is $\Delta =-\kappa /2$, with the results
\begin{align}
\Gamma _{\mathrm{opt}}^{^{\prime }}& =\frac{4G^{2}\omega _{m}(\kappa \omega
_{m}-4G^{2})}{2G^{4}+2\kappa \omega _{m}(\kappa \omega _{m}-4G^{2})}, \\
n_{\mathrm{opt}}^{^{\prime }}& =\frac{G^{2}}{2(\kappa \omega -4G^{2})}.
\end{align}%
The minimal value is obtained when $G=\sqrt{x\kappa \omega _{m}}$ with $x=%
\sqrt{n_{\mathrm{th}}/Q_{m}}/(4\sqrt{n_{\mathrm{th}}/Q_{m}}+1)$ and the minimum phonon number
is
\begin{equation}
n_{f,\mathrm{IS}}^{\min }=2\frac{n_{\mathrm{th}}}{Q_{m}}+\sqrt{\frac{n_{\mathrm{th}}}{Q_{m}}}.
\end{equation}

\section{Noise from the nonlinear mode}

\label{Sec6} Here we include the fluctuations of the second-order mode
induced by the nonlinear medium. For the quantum description of the
second-order nonlinear process, the Langevin equations read
\begin{align}
\dot{a_{1}}& =\left( i\Delta _{1}-\frac{\kappa _{1}}{2}\right) a_{1}+\nu
a_{1}^{\dag }a_{2}-ig_{1}a_{1}(b+b^{\dag })-\sqrt{\kappa _{1}}a_{1,\mathrm{in%
}}-\varepsilon _{1}, \\
\dot{a_{2}}& =\left( i\Delta _{2}-\frac{\kappa _{2}}{2}\right) a_{2}-\frac{%
\nu }{2}a_{1}^{2}-ig_{2}a_{2}(b+b^{\dag })-\sqrt{\kappa _{2}}a_{2,\mathrm{in}%
}-\varepsilon _{2}, \\
\dot{b}& =\left( -i\omega _{m}-\frac{\gamma }{2}\right) b-ig_{1}a_{1}^{\dag
}a_{1}-ig_{2}a_{2}^{\dag }a_{2}-\sqrt{\gamma }b_{\mathrm{in}},
\end{align}%
where $a_{1}$($a_{2}$) is the annihilation operator of the fundamental
(second-order) mode, $g_{1}$ ($g_{2}$) is the corresponding single-photon
optomechanical coupling strength, $\varepsilon _{1}$ ($\varepsilon _{2}$) is the
corresponding laser drive, $\kappa _{1}$ ($\kappa _{2}$) is the
corresponding total dissipation rate, and $a_{1,\mathrm{in}}$ ($a_{2,\mathrm{%
in}}$) is the corresponding noise operator, $b$ is the annihilation operator
of the mechanical mode, $\nu $ is the single photon $\chi ^{(2)}$
nonlinearity, and $b_{in}$ is the noise operator for the mechanical mode.
With the same linearization process by assuming $a_{1}=\alpha _{1}+\delta
a_{1}$, $a_{2}=\alpha _{2}+\delta a_{2}$, $b=\beta +\delta b$, we get the steady state satisfy

\begin{align}
0& =\left( i\Delta _{1}-\frac{\kappa _{1}}{2}\right) \alpha_{1}+\nu
\alpha_{1}^{\ast }\alpha_{2}-ig_{1}\alpha_{1}(\beta+\beta^{\ast })-\varepsilon _{1}, \\
0& =\left( i\Delta _{2}-\frac{\kappa _{2}}{2}\right) \alpha_{2}-\frac{%
\nu }{2}\alpha_{1}^{2}-ig_{2}\alpha_{2}(\beta+\beta^{\ast })-\varepsilon _{2}, \\
0& =\left( -i\omega _{m}-\frac{\gamma }{2}\right) \beta-ig_{1}\alpha_{1}^{\ast
}\alpha_{1}-ig_{2}\alpha_{2}^{\ast }\alpha_{2},
\end{align}

and the dynamics of the fluctuation fields yields
\begin{align}
\dot{\delta a_{1}}& =\left( i\Delta _{1,\mathrm{eff}}-\frac{\kappa _{1}}{2}\right)
\delta a_{1}+\nu \alpha _{2}\delta a_{1}^{\dag }+\nu \alpha _{1}^{\ast
}\delta a_{2}-ig_{1}\alpha _{1}(\delta b+\delta b^{\dag })-\sqrt{\kappa _{1}}%
a_{1,\mathrm{in}}, \\
\dot{\delta a_{2}}& =\left( i\Delta _{2,\mathrm{eff}}-\frac{\kappa _{2}}{2}\right)
\delta a_{2}-\nu \alpha _{1}\delta a_{1}-ig_{2}\alpha _{2}(\delta b+\delta
b^{\dag })-\sqrt{\kappa _{2}}a_{2,\mathrm{in}}, \\
\dot{\delta b}& =\left( -i\omega _{m}-\frac{\gamma }{2}\right) \delta
b-ig_{1}(\alpha _{1}^{\ast }\delta a_{1}+\alpha _{1}\delta a_{1}^{\dagger
})-ig_{2}(\alpha _{2}^{\ast }\delta a_{2}+\alpha _{2}\delta a_{2}^{\dagger
})-\sqrt{\gamma }b_{\mathrm{in}},
\end{align}%
where $\Delta _{i,\mathrm{eff}}=\Delta _{i}-g_{i}(\beta+\beta
^{\ast })$ ($i=1,2$). It can be seen that the effective nonlinear pumping $\varepsilon=\nu\alpha_2$ which depends on the circulation photons of the second-order optical mode in the cavity. By adiabatically eliminate the second-order mode,
we obtain
\begin{equation}
\delta a_{2}=\frac{1}{i\Delta _{2,\mathrm{eff}}-\frac{\kappa _{2}}{2}}[\nu
\alpha _{1}\delta a_{1}+ig_{2}\alpha _{2}(\delta b+\delta b^{\dag })+\sqrt{%
\kappa _{2}}a_{2,\mathrm{in}}].
\end{equation}%
First if $g_2=0$, it means that the radiation pressure of the second-order mode on mechanical oscillators is zero.
It happens for the standing wave of fundamental mode and second-order mode is different, hence a membrane putting in the cavity
with proper location will sustain a finite optomechanical coupling of fundamental mode but zero optomechanical coupling of second-order mode.
For this case, the added term of the Langevin equation compared with ones in main text with the reduced hamitionian is $\nu\alpha_1^{\ast}\delta a_2$ in the first line of the Langevin equations which yields

\begin{equation}
\nu\alpha_1^{\ast}\delta a_2=\frac{\nu\alpha_1^{\ast}}{i\Delta _{2,\mathrm{eff}}-\frac{\kappa _{2}}{2}}[\nu
\alpha _{1}\delta a_{1}+\sqrt{%
\kappa _{2}}a_{2,\mathrm{in}}].
\end{equation}

The first term will result an effective detuning and dissipation of fundamental mode $\Delta=\Delta_{1,eff}+\Delta_{2,eff}\nu^2|\alpha_1|^2/(\Delta_{2,eff}^2+\kappa_2^2/4)$ and $\kappa_{1,eff}=\kappa_1+\kappa_2\nu^2|\alpha_1|^2/(\Delta_{2,eff}^2+\kappa_2^2/4)$. The second term will result an additional vacuum noise which can be neglect compared with the effect induced by $-\sqrt{\kappa _{1}}a_{1,\mathrm{in}}$ when $\kappa_2\nu^2|\alpha_1|^2/(\Delta_{2,eff}^2+\kappa_2^2/4)\ll\kappa_1$, which is same the condition of neglecting the added dissipation of the fundamental optical mode. For the modification of detuning is not essential, hence the above amendments can be all safely neglected when $\Delta_{2,eff}\gg\sqrt{\kappa_2/\kappa_1}|\nu\alpha_1|$¡£

when the optomechanical coupling of the second-order optical mode is not vanish, there have three more added effect. One will result a modification of the optomechanical coupling between the fundamental optical mode and the mechanical mode which is not essential, too. The effective optomechanical coupling yields $G=g_1\alpha_1-g_2\nu\alpha_1^{\ast}\alpha_2/(i\Delta_{2,eff}-\kappa_2/2)$. The second effect is the mechanical squeezing with squeezing magnitude $\varepsilon_M=\frac{g^2|\alpha_2|^2\kappa_2}{\Delta_{2,eff}^2+\kappa^2/4}$, and it can be ignored when $\varepsilon_M\ll\omega_m$. Another effect is the induced fluctuation on the mechanical mode due to the radiation pressure of the second-order mode. This noise
finally leads to the additional thermal phonon occupancy, which can be
described by
\begin{equation}
b_{\mathrm{in}}^{\mathrm{add}}=\frac{-ig_{2}\sqrt{\kappa _{2}}}{\sqrt{\gamma
}(i\Delta _{2,\mathrm{eff}}-\frac{\kappa _{2}}{2})}[\alpha _{2}^{\ast }a_{2,%
\mathrm{in}}+\alpha _{2}a_{2,\mathrm{in}}^{\dagger }].
\end{equation}%
Then the added thermal phonon occupancy is
\begin{equation}
n_{\mathrm{in}}^{\mathrm{add}}=\frac{g_{2}^2|\alpha _{2}|^{2}\kappa _{2}}{%
\gamma (\Delta _{2,\mathrm{eff}}^{2}+\frac{\kappa _{2}^{2}}{4})}.
\end{equation}%
It is shown that the added noise can be merged into the environmental thermal noise with $b_{in}^{eff}=b_{in}+b_{in}^{add}$. Hence the demonstration is valid both in weak-coupling regime and strong coupling case. With the cooling process of intra-cavity squeezing, the added thermal phonon occupancy will result a final occupancy yields

\begin{equation}
n_{f}^{\textrm{add}}=\frac{g_{2}^2|\alpha _{2}|^{2}\kappa _{2}}{%
\omega_m (\Delta _{2,\mathrm{eff}}^{2}+\frac{\kappa _{2}^{2}}{4})}.
\end{equation}

When the nonlinear pumping is out of resonance from the cavity mode, i.e., $%
\Delta _{2,\mathrm{eff}}>>\sqrt{g_{2}^2|\alpha _{2}|^{2}\kappa _{2}/\omega_m}$, it can be seen that this is also the requirement of neglecting the mechanical squeezing effect induced by second-order optical mode. In conclusion, the system can be effectively described by the Linearized Hamitonian in the main tex when $\Delta_{2,eff}\gg\textrm{max}[\sqrt{g_{2}^2|\alpha _{2}|^{2}\kappa _{2}/\omega_m},\sqrt{\kappa_2/\kappa_1}|\nu\alpha_1|]$.

\section{power spectrum and squeezing properties}
We have introduced above that when the detuning of the second-order optical mode $\Delta_{2,eff}$ is large enough, the system can be effectively describe by the Linearized Hamitionian in the main tex and the output spectrum can be demonstrate from the following Langevin equations

\begin{align}
\dot{a_{1}}& =(i\Delta -\frac{\kappa }{2})a_{1}-iG(b_{1}+b_{1}^{\dag
})-2i\varepsilon a_{1}^{\dag }-\sqrt{\kappa}a_{\mathrm{in}%
}, \\
\dot{b_{1}}& =(-i\omega _{m}-\frac{\gamma }{2})b_{1}-i(Ga_{1}^{\dag
}+G^{\ast }a_{1}^{\dag })-\sqrt{\gamma }b_{\mathrm{in}},
\end{align}

Eliminating the mechanical mode, we get

\begin{align}
[\chi_c^{-1}(\omega)+H(\omega)]a_1(\omega)+[H(\omega)+2i\varepsilon]a_1^\dag(\omega)&=ig\sqrt{\gamma}(\chi_m^{-1} b_{in}(\omega)-\chi_m^{-1*}b_{in}^\dag(\omega))-\sqrt{\kappa}a_{in}(\omega)
\end{align}

where $\chi_c(\omega)=[-i(\omega+\Delta)+\kappa/2]^{-1}$, $\chi_m(\omega)=[-i(\omega-\omega_m)+\gamma/2]^{-1}$ are susceptibility of the optical mode and mechanical mode respectively. $H(\omega)=g^2(\chi_m^{-1}(\omega)-\chi_m^{-1*}(-\omega))$ and we can get $H^{*}(-\omega)=-H(\omega)$. It will results

\begin{equation}
a_1(\omega)=A_1(\omega)a_{in}(\omega)+A_2(\omega)a_{in}^\dag(\omega)+B_1(\omega)b_{in}(\omega)+B_2(\omega)b_{in}^\dag(\omega)
\end{equation}

where

\begin{align}
A_1(\omega)&=\frac{-\sqrt{\kappa}}{M}[\chi_m^{-1*}(-\omega)-H(\omega)]\\
A_2(\omega)&=\frac{-\sqrt{\kappa}}{M}[-2i\varepsilon-H(\omega)]\\
B_1(\omega)&=\frac{ig\sqrt{\gamma}\chi_m^{-1}(\omega)}{M}[\chi_c^{-1}(\omega)+2i\varepsilon+2H(\omega)]\\
B_2(\omega)&=\frac{-ig\sqrt{\gamma}\chi_m^{-1*}(-\omega)}{M}[\chi_c^{-1}(\omega)+2i\varepsilon+2H(\omega)]\\
\end{align}

and $M=[\chi_c^{-1}(\omega)+H(\omega)][\chi_c^{-1*}(-\omega)-H(\omega)]-[H+2i\varepsilon][-H-2i\varepsilon^{*}]$. From the above equations, we can obtain the power spectrum of the internal cavity quadrature $X=a_1e^{i\theta}+a_1^\dag e^{-i\theta}$. And the variance of the quadrature can be obtained by using $(\Delta X)^2=\int S_{XX}(\omega)d\omega$.

For the output spectrum, according to the input-output relationship $a_{1,out}(\omega)=a_{in}(\omega)-\sqrt{\kappa}a_1(\omega)$, we can get the output spectrum. For a particular quadrature $X=a_{1,out}e^{i\theta}+H.C.$, its power spectrum yields

\begin{equation}
S_{XX}(\omega)=S_{1}(\omega)+S_2(\omega)
\end{equation}

where

\begin{align}
S_1(\omega)&=|\sqrt{\kappa}A_2(-\omega)|^2+|\sqrt{\kappa}A_1(\omega)|^2+\kappa(|B_1(\omega)|^2+|B_2(-\omega)|^2)(n_{th}+1)+\kappa(|B_1(-\omega)|^2+|B_2(\omega)|^2)n_{th}\\
S_2(\omega)&=2 Re[e^{-2i\theta}(1-\sqrt{\kappa}A_1(\omega))\sqrt{\kappa}A_2(-\omega)+\kappa e^{-2i\theta}B_1(\omega)B_2(-\omega)(n_{th}+1)+\kappa e^{-2i\theta}B_1(-\omega)B_2(\omega)n_{th}]
\end{align}

Hence the squeezing phase at different frequency yields $\theta_{opt}=(\phi-\pi)/2$, with $\phi=Arg[(1-\sqrt{\kappa}A_1(\omega))\sqrt{\kappa}A_2(-\omega)+\kappa B_1(\omega)B_2(-\omega)(n_{th}+1)+\kappa B_1(-\omega)B_2(\omega)n_{th}]$. And the squeezing magnitude is

\begin{equation}
r(\omega)=Log\{Min[S_{XX}(\omega)]\}=Log\{S_1-|(1-\sqrt{\kappa}A_1(\omega))\sqrt{\kappa}A_2(-\omega)+\kappa B_1(\omega)B_2(-\omega)(n_{th}+1)+\kappa B_1(-\omega)B_2(\omega)n_{th}|\}
\end{equation}

\end{document}